\documentclass[12pt]{article}

\newcommand{\sect}[1]{\setcounter{equation}{0}\section{#1}}

\global\arraycolsep=1pt
\def\fnote#1#2{\begingroup\def\thefootnote{#1}\footnote{#2}\addtocounter
{footnote}{-1}\endgroup}

\usepackage{amsbsy,amssymb,latexsym,epsfig}

\newcommand{\beq}{\begin{equation}}
\newcommand{\eeq}{\end{equation}}
\newcommand{\beqa}{\begin{eqnarray}}
\newcommand{\eeqa}{\end{eqnarray}}
\newcommand{\bR}{{\mathbb R}}

\newcommand{\CL}{{\mathcal L}}

\def\fnote#1#2{\begingroup\def\thefootnote{#1}\footnote{#2}\addtocounter
{footnote}{-1}\endgroup}

\usepackage{amsbsy,amssymb,latexsym}

\begin{document}
\begin{flushright}
OCU-PHYS 226 \\
hep-th/0502182

\end{flushright}
\vspace{12mm}

\begin{center}
{\bf\Large
Notes on 
Five-dimensional Kerr Black Holes
}

\vspace{15mm}
Makoto Sakaguchi\fnote{$\star$}{
\texttt{msakaguc@sci.osaka-cu.ac.jp}
}
and
Yukinori Yasui\fnote{$\ast$}{
\texttt{yasui@sci.osaka-cu.ac.jp}
}
\vspace{10mm}

\textit{
${{}^\star}$
Osaka City University
Advanced Mathematical Institute (OCAMI)
}
\vspace{2mm}

${}^\ast$
\textit{
Department of Mathematics and Physics,
Graduate School of Science,\\
Osaka City University
}

\vspace{3mm}

\textit{
Sumiyoshi,
Osaka 558-8585, JAPAN}
\end{center}
\vspace{8mm}

\begin{abstract}
The geometry of five-dimensional Kerr black holes is
discussed based on geodesics and Weyl curvatures.
Kerr-Star space, Star-Kerr space
and Kruskal space are naturally
introduced by using special null geodesics.
We show that the geodesics of AdS Kerr black hole are integrable,
which generalizes the result of Frolov and Stojkovic.
We also show that five-dimensional AdS Kerr black holes are
isospectrum deformations of Ricci-flat
Kerr black holes in the sense that
the eigenvalues of the Weyl curvature
are preserved.
\end{abstract}
\newpage

\sect{Introduction}

Black holes have attracted renewed interests
in the recent developments of
string theory and Riemannian geometry,
such as
the AdS/CFT correspondence \cite{AdS/CFT}
and a certain relation between compact Einstein manifolds
and black holes
\cite{Page}
\cite{HSY}
\cite{Gibbons}
\cite{HSY2}.
These developments
motivate us to
study the geometry of black holes
especially in four-, five- and seven-dimensions
among
higher dimensional black holes.

This paper is
devoted to
a first step to investigate the global geometry 
of five-dimensional Kerr black holes constructed by 
Myers and Perry \cite{MP},
and Hawking et al \cite{Hawking}.
For four-dimensional Kerr black holes,
a valuable textbook \cite{O'Neill}
written by
B. O'Neill
from mathematical point of view
has been published.
In the textbook,
the differential geometry based on special null geodesics
(principal null geodesics)
were fully analyzed.
Following the textbook
we try
to generalize the analysis to five-dimensional
black holes.
In this paper
we do not stick to mathematical completeness,
but develop some key points
of the geometry;
integrability of geodesics
and curvature property.
These points were
studied in
the previous works
\cite{Frolov:2002xf} and \cite{DeSmet:2003bt},
however, our method is different from them.
In addition,
AdS black holes have not been discussed in the textbook.
We show that five-dimensional AdS Kerr black holes are
isospectrum deformations of Ricci-flat
Kerr black holes in the sense that
the eigenvalues of the Weyl curvature
are preserved.

This paper is organized as follows.
In the next section,
we examine the curvature property
of five-dimensional black holes.
We show that a linear map on two-forms
constructed from Riemannian curvature 
is diagonalizable,
and derive the eigenvalues and the degeneracy.
In section 3,
we examine the integrability of
geodesics
based on the Euler-Lagrange equations.
The section 4 is devoted to the global analysis of
five-dimensional Kerr black holes.
In the last section,
we generalize the analysis
to the case with cosmological constant.
We show the integrability of geodesics
and examine the curvature property.
In appendix A, the analysis on the four-dimensional
AdS Kerr black hole is given.

While this paper was in preparation
we received \cite{KL}
which overlaps with the result 
on the integrability of geodesics
on the five-dimensional AdS Kerr black hole.
However their result has been
obtained using a different approach.


\sect{5-dimensional Kerr Black Holes}
Let us write the Ricci flat metric for
the 5-dimensional Kerr black hole \cite{MP}
in the Boyer-Lindquist coordinates
$(t,\phi,\psi,r,\theta)$;
\begin{eqnarray}
g_{BH}&=&
-dt^2
+\frac{\rho^2}{\Delta_r}dr^2
+\rho^2 g_{\rm{S}^3}
+(a^2-b^2)(\sin^2\theta d\phi)^2
+(b^2-a^2)(\cos^2\theta d\psi)^2
\nonumber\\&&
+\frac{2m}{\rho^2}
(dt-a\sin^2\theta d\phi-b\cos^2\theta d\psi)^2~,
\label{metric:g_BH}
\end{eqnarray}
where
\begin{eqnarray}
\rho^2&=&
r^2
+a^2\cos^2\theta
+b^2\sin^2\theta~,\\
\Delta_r&=&
\frac{(r^2+a^2)(r^2+b^2)}{r^2}-2m~.
\end{eqnarray}
The non-negative parameters $a$, $b$ and $m$
correspond to
angular momenta and mass, respectively.
The metric $g_{\rm{S}^3}$
is given by
\begin{eqnarray}
g_{\rm{S}^3}&=&
d\theta^2+\sin^2\theta d\phi^2
+\cos^2\theta d\psi^2
\label{matric:S3}
\end{eqnarray}
with the ranges $0\le \theta\le \pi/2$
and $0\le \phi,\psi \le 2\pi$.
If we introduce an orthonormal coordinate $(x,y,z,w)$;
\begin{eqnarray}
x=\sin\theta\cos\phi~,~~~
y=\sin\theta\sin\phi~,~~~
z=\cos\theta\cos\psi~,~~~
w=\cos\theta\sin\psi~,
\end{eqnarray}
then
\begin{eqnarray}
g_{\rm{S}^3}&=&dx^2+dy^2+dz^2+dw^2~,
\end{eqnarray}
which represents the standard metric on a three-sphere S$^3$.
Also, the one-forms $\sin^2\theta d\phi$ and $\cos^2\theta d\psi$
in the metric (\ref{matric:S3}) are written as
\begin{eqnarray}
\sin^2\theta d\phi =xdy-ydx~,~~~
\cos^2\theta d\psi =zdw-wdz~,
\end{eqnarray}
and hence these are well-defined as one-forms on S$^3$.
Thus $g_{BH}$ can be regarded as a metric on the Lorentzian space
\begin{eqnarray}
M=\bR_+^2\times {\rm S}^3 -H~,
\end{eqnarray}
where $\bR_+^2 =\{(t,r)~|~t\in\bR, r\in\bR^+\}$.
The horizon $H\simeq \bR\times {\rm S}^3$
defined by $\Delta_r=0$
is removed from the space
since $\rho^2/\Delta_r\to\infty$ on $H$.
The equation $\Delta_r=0$ has two positive real roots $r=r_\pm$,
\begin{eqnarray}
r_\pm^2&=&
\frac{1}{2}\left(
2m-a^2-b^2\pm \sqrt{\Big(2m-(a-b)^2\Big)\Big(2m-(a+b)^2\Big)}
\right)
\end{eqnarray}
in the region $m>(a+b)^2/2$.
We have three open sets I, II and III (called Boyer-Lindquist blocks)
of M:
\begin{eqnarray}
{\rm I}: r_+< r~,~~~
{\rm II}: r_-<r< r_+~,~~~
{\rm III}: 0<r<r_-~.
\end{eqnarray}
It should be noticed that three-dimensional submanifolds
$M|_{\theta=0}$ and $M|_{\theta=\pi/2}$
are time-like totally geodesic,
while two-dimensional submanifold $M|_{t,\phi,\psi={\rm const}}$
is space-like totally geodesic.
This can be seen from calculations of the Christoffel symbol;
$\Gamma^\theta_{\mu\nu}\propto \sin 2\theta$
and $\Gamma^i_{\mu\nu}=0$ ($i=t,\phi,\psi$),
where $\mu$ and $\nu$ run the tangential indices to the submanifolds.

Let us calculate the Riemannian curvature of the black hole metric.
It is convenient to use the following orthonormal
frame $\{e_a\}(a=1,2,\dots,5)$ \cite{DeSmet:2003bt}
\cite{AF}:
\begin{eqnarray}
e_1&=&
\frac{1}{r^2\sqrt{\Delta_r\varepsilon}\rho} \left(
\Sigma_a^2\Sigma_b^2\frac{\partial}{\partial t}
+a\Sigma_b^2\frac{\partial}{\partial \phi}
+b\Sigma_a^2\frac{\partial}{\partial \psi} \right)~,
\label{e1:MP}\\
e_2&=&
\frac{1}{r\rho\sin\theta}\left[
\Sigma_b \left(\frac{\partial}{\partial \phi}
+a\sin^2\theta\frac{\partial}{\partial t} \right)
+\frac{b\Sigma_a}{\Sigma_a\Sigma_b+r\rho}\left(
a\sin^2\theta\frac{\partial}{\partial \psi}
-b\cos^2\theta\frac{\partial}{\partial \phi} \right)
\right]~,~~\\
e_3&=&
\frac{1}{r\rho\cos\theta}\left[
\Sigma_a \left(\frac{\partial}{\partial \psi}
+b\cos^2\theta\frac{\partial}{\partial t} \right)
-\frac{a\Sigma_b}{\Sigma_a\Sigma_b+r\rho}
\left(a\sin^2\theta\frac{\partial}{\partial \psi}
-b\cos^2\theta\frac{\partial}{\partial \phi} \right)
\right]~,~~~~\\
e_4&=&
\frac{\sqrt{\Delta_r\varepsilon}}{\rho}
\frac{\partial}{\partial r}~,\\
e_5&=&
\frac{1}{\rho}\frac{\partial}{\partial \theta}~,
\label{e5:MP}
\end{eqnarray}
where
$\Sigma_a=\sqrt{r^2+a^2}$ and
$\Sigma_b=\sqrt{r^2+b^2}$. The inner product is given by
$\langle e_a ,e_b\rangle_{g_{BH}}=\eta_{ab}$ with
$\eta_{ab}={\rm diag}(-\varepsilon,+1,+1,\varepsilon,+1)$.
The time-like vector is $e_1$ in I $\cup$ III,
while $e_4$ in II,
so that we choose 
$\varepsilon= 1$ in I $\cup$ III and $\varepsilon=- 1$ in II.

The Riemannian curvature
$R^{ab}{}_{cd}$ may be regarded as a linear map (curvature transformation)
on two-forms
$\Lambda^2(M)$. 
Then, the matrix
$R:\Lambda^2(M) \rightarrow \Lambda^2(M) $ is of
the form $R^{ab}{}_{cd}=\frac{2m r^2}{\rho^6}\hat R^{ab}{}_{cd}$.
Explicitly, non-zero components of $\hat R^{ab}{}_{cd}$ are given as
\begin{eqnarray}
  \begin{array}{c|ccccc}
(a,b)\setminus (c,d)      &(1,2)     &(1,3)     &(1,4)    &(1,5)    &(2,3)    
          \\\hline
(1,2)       &-1+{\rm\bf I}^2-{\rm\bf J}^2    ~~&-2{\rm\bf I}{\rm\bf J}    &    &    &    
\\
(1,3)       &-2{\rm\bf I}{\rm\bf J}    ~~&-1-{\rm\bf I}^2-{\rm\bf J}^2    &    &    &    
\\
(1,4)       &    &    &3-{\rm\bf I}^2-{\rm\bf J}^2    &    &    
\\
(1,5)       &    &    &    & -1+{\rm\bf I}^2+{\rm\bf J}^2   &    
\\
(2,3)       &    &    &    &    &1+{\rm\bf I}^2+{\rm\bf J}^2    
\\
(2,4)       &    &    &    &-{2}{\varepsilon}{\rm\bf I}    &    
\\
(2,5)       &    &    &-4{\rm\bf I}    &    &    
\\
(3,4)       &    &    &    &{2}{\varepsilon}{\rm\bf J}    &    
\\
(3,5)       &    &    & 4{\rm\bf J}   &    &    
\\
(4,5)       &-{2}{\varepsilon}{\rm\bf I}    &{2}{\varepsilon}{\rm\bf J}    &    &    &    
\\
  \end{array}\nonumber\\
  \begin{array}{c|ccccc}
(a,b)\setminus (c,d)          
       &(2,4)     &(2,5)     & (3,4)   &(3,5)     &(4,5)    \\\hline
(1,2)           
&    &    &    &    &2\varepsilon {\rm\bf I}    \\
(1,3)           
&    &    &    &    &-2\varepsilon {\rm\bf J}    \\
(1,4)           
&    & 4{\rm\bf I}   &    &-4{\rm\bf J}    &    \\
(1,5)           
& 2\varepsilon {\rm\bf I}   &    & -2\varepsilon {\rm\bf J}   &    &    \\
(2,3)           
&    &    &    &    &    \\
(2,4)          
&-1+{\rm\bf I}^2-{\rm\bf J}^2    &    &-2{\rm\bf I}{\rm\bf J}    &    &    \\
(2,5)           
&    & 1-3{\rm\bf I}^2+{\rm\bf J}^2   &    &4{\rm\bf I}{\rm\bf J}    &    \\
(3,4)           
&-2{\rm\bf I}{\rm\bf J}    &    &-1-{\rm\bf I}^2+{\rm\bf J}^2    &    &    \\
(3,5)           
&    & 4{\rm\bf I}{\rm\bf J}   &    &1+{\rm\bf I}^2-3{\rm\bf J}^2    &    \\
(4,5)           
&    &    &    &   &-1+{\rm\bf I}^2+{\rm\bf J}^2     \\
  \end{array}\nonumber\\
\label{matrix}
\end{eqnarray}
where
\begin{eqnarray}
{\rm\bf I}&=&
\frac{a\cos\theta}{r}\,
\frac{\Sigma_ar+\Sigma_b\rho}{\Sigma_a\Sigma_b+r\rho}~,~~~~
{\rm\bf J}=
\frac{b\sin\theta}{r}\,
\frac{\Sigma_br+\Sigma_a\rho}{\Sigma_a\Sigma_b+r\rho}~.
\label{I and J}
\end{eqnarray}

We find that the $10\times 10$ matrix $R^{ab}{}_{cd}$
is diagonalizable
and
the eigenvalues
depend on the combination
${\rm\bf K}^2\equiv {\rm\bf I}^2+{\rm\bf J}^2$
as
\begin{eqnarray}
2&~:~& \lambda_1=
\frac{2 m r^2}{\rho^6}(1+{\rm\bf K}^2)
=\frac{2m}{\rho^4} ~, 
\label{lambda 1}         \\
2&:& \lambda_2=
-\frac{2 m r^2}{\rho^6}(1+{\rm\bf K}^2)
=-\frac{2m}{\rho^4}      ~,     \\
2&:& \lambda_3=\frac{2 m r^2}{\rho^6}
(-1+{\rm\bf K}^2+2 \sqrt{-{\rm\bf K}^2})\nonumber\\
&&~~~
=\frac{2m}{\rho^6}\left(
\rho^2-2r^2
+2ir\sqrt{{
\rho^2-r^2
}
}
\right)       ~,     \\
2&:& \lambda_4=\frac{2 m r^2}{\rho^6}
(-1+{\rm\bf K}^2-2 \sqrt{-{\rm\bf K}^2})\nonumber\\ 
&&~~~
=\frac{2m}{\rho^6}\left(
\rho^2-2r^2
-2ir
\sqrt{
{
\rho^2-r^2
}
}
\right)         ~,     
\label{lambda 4}\\
1&:& \lambda_5=\frac{2 m r^2}{\rho^6}
(2-2{\rm\bf K}^2
+\sqrt{1-14{\rm\bf K}^2
+{\rm\bf K}^4}) \nonumber\\
&&~~~=\frac{2m}{\rho^6}\Biggl(
-2\rho^2+4r^2
+\sqrt{
{(r^2-(7+4\sqrt{3})(
\rho^2-r^2
))
(r^2-(7-4\sqrt{3})(
\rho^2-r^2
))}
}
\Bigr)    ~,    \nonumber\\&&\\
1&:& \lambda_6=\frac{2 m r^2}{\rho^6}
(2-2{\rm\bf K}^2
-\sqrt{1-14{\rm\bf K}^2
+{\rm\bf K}^4})\nonumber\\ 
&&~~~
=\frac{2m}{\rho^6}\Biggl(
-2\rho^2+4r^2
-\sqrt{
{(r^2-(7+4\sqrt{3})(
\rho^2-r^2
))
(r^2-(7-4\sqrt{3})(
\rho^2-r^2
))}
}
\Bigr) ~, 
\nonumber\\&&\label{lambda 6}
\end{eqnarray}
where the number in the left hand side represents the degeneracy.

The curvature invariants ${\rm tr} (R^n)$ ($n=1,2,3,...$)
can be simply written as
\begin{eqnarray}
{\rm tr} (R^n)=
2\sum_{i=1}^{4}(\lambda_i)^n+(\lambda_5)^n+(\lambda_6)^n=
\left(\frac{2m}{\rho^{6}}\right)^nP_n(r^2,\theta)~,
\end{eqnarray}
where $P_n$ is an $n$-polynomial
with respect to $r^2$.
For example, we have
\begin{eqnarray}
P_1&=&
0
~,\\
P_2&=&
6r^4
(3{\rm\bf K}^2-2)
({\rm\bf K}^2-3)
~,\\
P_3&=&
-24r^6
({\rm\bf K}^2-1)
\left(
{\rm\bf K}^4
-6{\rm\bf K}^2
+1
\right)
~.
\end{eqnarray}
The invariants are finite
for $ab\neq 0$.
However,
when we extend the square of the radial coordinate $r^2$
to a negative region,
a singularity
appears at $\rho^2=0$,
i.e.
$r^2=-a^2\cos^2\theta -b^2\sin^2\theta$.

For the two-dimensional subspace $\Pi_{ab}$
spanned by  $\{e_a, e_b\}$,
the sectional curvature is defined by
$K(\Pi_{ab})=\eta_{bb}R_{abab}$.
We find from (\ref{matrix}) and (\ref{lambda 1})-(\ref{lambda 4})
the following relation between
the sectional curvature and the eigenvalue with
degeneracy 2:
\begin{eqnarray}
&&K(\Pi_{15})=K(\Pi_{45})=-{\rm Re}\lambda_3=-{\rm Re}\lambda_4~,\\
&&K(\Pi_{23})=\lambda_1=-\lambda_2~.
\end{eqnarray}

\sect{Geodesics}
In this section, we examine geodesics
and show the integrability of them.
The integrability of geodesics has been shown
in \cite{Frolov:2002xf} by using Hamilton-Jacobi formulation,
while we work in the Euler-Lagrange formulation.

The geodesic equations are given by
the Euler-Lagrange equations.
From the metric (\ref{metric:g_BH})
we obtain a Lagrangian
\begin{eqnarray}
\CL=\frac{1}{2}g_{\alpha\beta}\frac{dx^\alpha}{ds}\frac{dx^\beta}{ds}~,~~~
(\alpha, \beta =t,\phi,\psi,r,\theta)~,
\end{eqnarray}
where
\begin{eqnarray}
g_{tt}&=&
-1+\frac{2m}{\rho^2}~,\label{metric top}\\
g_{t\phi}&=&
-\frac{2am\sin^2\theta}{\rho^2}~,\\
g_{t\psi}&=&
-\frac{2bm\cos^2\theta}{\rho^2}~,\\
g_{\phi\psi}&=&
\frac{2abm\cos^2\theta\sin^2\theta}{\rho^2}~,
\label{p:top}\\
g_{\phi\phi}&=&
\sin^2\theta\left(
r^2+a^2+\frac{2ma^2\sin^2\theta}{\rho^2}
\right)~,\\
g_{\psi\psi}&=&
\cos^2\theta\left(
r^2+b^2+\frac{2mb^2\cos^2\theta}{\rho^2}
\right)~,
\label{p:bottom}\\
g_{rr}&=&
\frac{\rho^2}{\Delta_r}~,\\
g_{\theta\theta}&=&
\rho^2~.\label{metric bottom}
\end{eqnarray}
The metric has three Killing vector fields
$\{\frac{\partial}{\partial t},
\frac{\partial}{\partial \phi},
\frac{\partial}{\partial \psi}\}$,
and hence there are three conservations $L_t$, $L_\phi$
and $L_\psi$.
In addition, we have a ``total energy'' $Q$
corresponding to the Hamiltonian.
Let us write a geodesic as
$\gamma(s)=(t(s), \phi(s),\psi(s), r(s),\theta(s))$.
Then the conservations above are given by
\begin{eqnarray}
L_t= \langle \gamma', \frac{\partial}{\partial t}\rangle~,~~~
L_\phi= \langle \gamma', \frac{\partial}{\partial \phi}\rangle~,~~~
L_\psi= \langle \gamma', \frac{\partial}{\partial \psi}\rangle~,~~~
Q=\langle \gamma', \gamma' \rangle~,
\end{eqnarray}
where
$\gamma'=\frac{d\gamma}{ds}$,
and $\langle\cdot ,\cdot \rangle$
represents an inner product with respect to the black hoke metric.

In addition,
there exists
another conservation
generalizing the Carter constant in the four-dimensional Kerr black hole
\cite{Carter}.
\medskip

\noindent
{\bf \underline{Theorem 1.}}~(see also \cite{Frolov:2002xf})\\
{\it 
There exists a constant $K$ for a geodesic
$\gamma$ satisfying
\begin{eqnarray}
&&\rho^4\left(\frac{dr}{ds}\right)^2+R(r)=0~,\\
&&\rho^4\left(\frac{d\theta}{ds}\right)^2+\Theta(\theta)=0~,
\end{eqnarray}
where
\begin{eqnarray}
\Theta&=&
(a^2\sin^2\theta+b^2\cos^2\theta)L_t^2
+\frac{L_\phi^2}{\sin^2\theta}
+\frac{L_\psi^2}{\cos^2\theta}
-(a^2\cos^2\theta+b^2\sin^2\theta)Q
-K~,
\label{Theta:l=0}\\
R&=&
\frac{R_{-2}}{r^2}
+R_0
+r^2R_2
+r^4R_4~.
\label{R:l=0}
\end{eqnarray}
The coefficients are explicitly given by
\begin{eqnarray}
R_{4}&=&-Q-L_t^2~,\\
R_2&=&
-2(a^2+b^2)L_t^2
+Q(a^2+b^2-2m)
+K
~,\\
R_0&=&
-(a^4+b^4+3a^2b^2)L_t^2
-(a^2-b^2)(L_\phi^2-L_\psi^2)
-4mL_t(aL_\phi+bL_\psi)
\nonumber\\&&
-a^2b^2Q+(a^2+b^2-2m)K
~,\\
R_{-2}&=&
-a^2b^2(a^2+b^2+2m)L_t^2
-b^2(a^2-b^2+2m)L_\phi^2
-a^2(b^2-a^2+2m)L_\psi^2
\nonumber\\&&
-4abm\left(
L_t(bL_\phi+aL_\psi)
+L_\phi L_\psi
\right)
+K a^2b^2
~.
\end{eqnarray}
}
\medskip

\noindent
\textbf{\underline{Proof}}

For simplicity we prove this theorem
in a special case;
the
totally geodesic submanifold $N\equiv M|_{t,\phi,\psi={\rm const.}}$.
The metric restricted to $N$
is given by $g_N=\frac{\rho^2}{\Delta_r}dr^2+\rho^2 d\theta^2$.
The Euler-Lagrange equations are
\begin{eqnarray}
\frac{d}{ds}\left(\frac{\rho^2}{\Delta_r}\frac{dr}{ds}\right)&=&
\frac{1}{2}\left[
\frac{\partial}{\partial r}\left(\frac{\rho^2}{\Delta_r}\right)
\left(\frac{dr}{ds}\right)^2
+\frac{\partial\rho^2}{\partial r}\left(\frac{d\theta}{ds}\right)^2
\right]~,\label{geodesic eqn 1}\\
\frac{d}{ds}\left(\rho^2\frac{d\theta}{ds}\right)&=&
\frac{1}{2}\left[
\frac{\partial}{\partial \theta}\left(\frac{\rho^2}{\Delta_r}\right)
\left(\frac{dr}{ds}\right)^2
+\frac{\partial\rho^2}{\partial \theta}\left(\frac{d\theta}{ds}\right)^2
\right]~.\label{geodesic eqn 2}
\end{eqnarray}
From (\ref{geodesic eqn 2}), we have
\begin{eqnarray}
\frac{d}{ds}\left(\rho^2\frac{d\theta}{ds}\right)=
-\frac{1}{2}(a^2-b^2)\sin 2\theta
\left(\frac{1}{\Delta_r}\left(\frac{dr}{ds}\right)^2
+\left(\frac{d\theta}{ds}\right)^2
\right)~.
\label{geodesic eqn: proof}
\end{eqnarray}
By using the conservation
\begin{eqnarray}
Q=\langle\gamma',\gamma'\rangle_{g_N}
=\rho^2\left(
\frac{1}{\Delta_r}\left(\frac{dr}{ds}\right)^2
+\left(\frac{d\theta}{ds}\right)^2
\right)~,
\end{eqnarray}
the equation (\ref{geodesic eqn: proof})
can be transformed to
\begin{eqnarray}
\frac{d}{ds}\left[
\left(\rho^2\frac{d\theta}{ds}\right)^2
-\frac{1}{2}Q(a^2-b^2)\cos 2\theta
\right]=0~,
\end{eqnarray}
which yields
\begin{eqnarray}
\rho^4\left(\frac{d\theta}{ds}\right)^2=
\frac{1}{2}Q(a^2-b^2)\cos 2\theta +K
\label{kappa-1}
\end{eqnarray}
where $K$ is an integration constant.
It is easy to show that
\begin{eqnarray}
\rho^4 \left(\frac{dr}{ds}\right)^2=
\Delta_r\left(Q(r^2+\frac{a^2+b^2}{2})-K\right)~.
\label{kappa-2}
\end{eqnarray}
The equations (\ref{kappa-1}) and (\ref{kappa-2})
give the first-order geodesic equations on $N$,
which are identical
 to the theorem specialized to the case $L_t=L_\phi=L_\psi=0$
 with a trivial constant shift of $K$.
Using similar arguments we can prove the theorem
in the general case, although
the calculation is complicated.
\hfill $\Box$

\sect{Maximal Extension of Kerr spacetime} 
Let us consider
special geodesics
$\gamma_{\{{{\rm out} \atop {\rm in}}}(s)$ with
constants (where $\theta$ is an arbitrary constant)
\begin{eqnarray}
&&Q=0~,~~~ 
L_t=-1~,~~~ 
L_\phi=a\sin^2\theta~,~~~ 
L_\psi=b\cos^2\theta~,
\nonumber\\&&
K=2(a^2\sin^2\theta+b^2\cos^2\theta)
\end{eqnarray}
for which $\Theta=0$ and $R(r)=-\rho^4$~:
\begin{eqnarray}
\gamma'_{\{{{\rm out} \atop {\rm in}}}&=&
\pm\frac{\partial}{\partial r}
+\frac{1}{r^2\Delta_r}V(r)~,
\label{special geodesic}
\end{eqnarray}
where
\begin{eqnarray}
V&=&
(r^2+a^2)(r^2+b^2)\frac{\partial}{\partial t}
+a(r^2+b^2)\frac{\partial}{\partial \phi}
+b(r^2+a^2)\frac{\partial}{\partial \psi}~.
\end{eqnarray}
For the limit $r\to \infty$, 
$\gamma'_{\{{{\rm out} \atop {\rm in}}}\to \pm\frac{\partial}{\partial r}
+\frac{\partial}{\partial t}$,
which represents an outgoing (or ingoing, respectively) null
geodesic in the Boyer-Lindquist block I.
The direction of the geodesic $\gamma_{\{{{\rm out}\atop {\rm in}}}$
coincides with 
$e_1\pm e_4$,
\begin{eqnarray}
\pm\frac{\partial}{\partial r}+\frac{1}{r\Delta_r}V
=\frac{\rho}{\sqrt{\Delta_r\varepsilon}}(e_1\pm e_4)
~.
\end{eqnarray}
It should be noticed that the ingoing null geodesic $\gamma_{\rm in}$
coincides with the Kerr-Schild null geodesic
given in \cite{MP}.

We introduce a new coordinate (Kerr-Star coordinate)
defined by
\begin{eqnarray}
t^*&=&
t+T(r)~,\\
\phi^*&=&
\phi+A(r)~,\\
\psi^*&=&
\psi+B(r)~,
\end{eqnarray}
together with $r^*=r$ and $\theta^*=\theta$, where
\begin{eqnarray}
T(r)&=&
\int \frac{(r^2+a^2)(r^2+b^2)}{r^2\Delta_r}dr~,
\label{T}\\
A(r)&=&
\int \frac{a(r^2+b^2)}{r^2\Delta_r}dr~,
\label{A}\\
B(r)&=&
\int \frac{b(r^2+a^2)}{r^2\Delta_r}dr~.
\label{B}
\end{eqnarray}
Then, the null geodesics are written as \fnote{$\dagger$}{
Note that 
$\frac{\partial}{\partial r^*}=\frac{\partial}{\partial r}-\frac{1}{r^2\Delta_r}V$,
although $r^*=r$.
We have also 
$
\frac{\partial}{\partial t^*}=\frac{\partial}{\partial t},
\frac{\partial}{\partial \phi^*}=\frac{\partial}{\partial \phi},
\frac{\partial}{\partial \psi^*}=\frac{\partial}{\partial \psi}
$.
}
\begin{eqnarray}
\gamma'_{\rm in}&=&
-\frac{\partial}{\partial r^*}~,
\label{geodesic in}\\
\gamma'_{\rm out}&=&
\frac{\partial}{\partial r^*}
+\frac{2}{r^2\Delta_r}V~.\label{geodesic out}
\end{eqnarray}
The metric in the Kerr-Star coordinate is given
by
\begin{eqnarray}
g_{BH}&=&
\sum_{i,j=t,\phi,\psi}
g_{ij}dx^{*i}dx^{*j}
+g_{\theta\theta}d\theta^2
+2dt^*dr
-2a\sin^2\theta drd\phi^*
-2b\cos^2\theta drd\psi^*
~,
\label{metric:Kerr-Star}
\end{eqnarray}
where the components $g_{ij}$ and $g_{\theta\theta}$
are same form as ones written by the Boyer-Lindquist coordinate
(see (\ref{metric:g_BH}) or (\ref{metric top})-(\ref{metric bottom})).

\begin{figure}
   \parbox{\textwidth}{
   \begin{center}
      \psfig{file=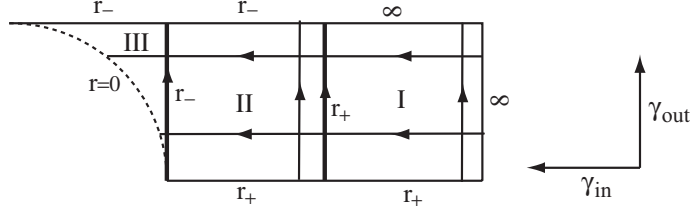,height=30mm}
   \end{center}
   \caption{\textit{Kerr-Star space $M^*$.}~~
   {\small 
   Horizontal lines crossing the horizons represent ingoing null geodesics,
   while vertical lines outgoing null geodesics.
   Ingoing null geodesics run from $r=\infty$ to $r=0$.
   Outgoing null geodesics in the Boyer-Lindquist block I(III)
   run from $r=r_+$ ($r=0$) to $r=\infty$ ($r=r_-$); the radial coordinate increases
   since $\Delta_r>0$.
   In the block II, outgoing null geodesics run from $r=r_+$
   to $r=r_-$;
   the radial coordinate decreases since $\Delta_r<0$.
}   }
\label{K*}
}
\end{figure}

The dangerous term at the horizon, $g_{rr}\propto1/\Delta_r$,
is absent, 
and so $g_{BH}$ is extended to the metric $g_{BH}^*$
on the space $M^*=\bR^2_+\times$S$^3$
(see figure \ref{K*}).
The outgoing null geodesic $\gamma_{\rm out}$
is not defined on $H$.
However, multiplying the factor $r^2\Delta_r/2$ with
the right hand side in (\ref{geodesic out}),
we have a vector field
\begin{eqnarray}
X^*&=&
\frac{r^2\Delta_r}{2}\frac{\partial}{\partial r^*}
+V~,\label{X*}
\end{eqnarray}
which is defined on $M^*$ including $H$.
Thus, one may understand the outgoing null geodesic $\gamma_{\rm out}$
as an integral curve of $X^*$.
On the surface $r=$constant,
the determinant of $g^*_{BH}|_{r={\rm const.}}$
is calculated as 
$-\rho^2r^2\Delta_r\sin^2\theta\cos^2\theta$,
so that the restricted metric is degenerate if $\Delta_r=0$,
i.e.
the horizons $H_{\pm}=H|_{r=r_\pm}$
are null hypersurfaces.
At $H_\pm$,
$X^*$ in (\ref{X*}) reduces to
the Killing vector fields
\begin{eqnarray}
V_\pm&=&V|_{r=r_\pm}=
(r_\pm^2+a^2)(r_{\pm}^2+b^2)\frac{\partial}{\partial t}
+a(r_\pm^2+b^2)\frac{\partial}{\partial \phi}
+b(r_\pm^2+a^2)\frac{\partial}{\partial \psi}~.
\end{eqnarray}
The vector fields $V_\pm$ are tangential to $H_\pm$
and also perpendicular to
$H_{\pm}$.
The integral curves of $V_\pm$ generate the horizons,
which become totally geodesic null hypersurfaces in $M^*$.

\begin{figure}
   \parbox{\textwidth}{
   \begin{center}
      \psfig{file=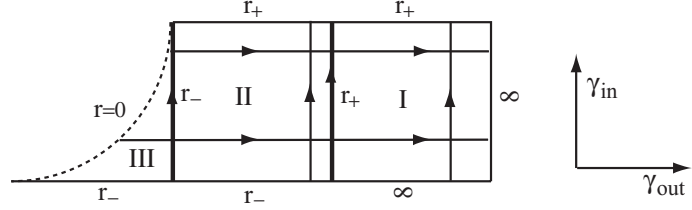,height=30mm}
   \end{center}
   \caption{ \textit{Star-Kerr space $^*\!M$.}~~
   {\small 
   Horizontal lines crossing the horizons
   represent outgoing null geodesics, while
   vertical lines ingoing null geodesics.
   Star-Kerr space is opposite
   to the Kerr-Star space in these respects.   
   }
   }
\label{*K}
}
\end{figure}

If we introduce a coordinate (Star-Kerr coordinate),
\begin{eqnarray}
^*\!t=t-T(r)~,~~~
^*\!\phi=\phi-A(r)~,~~~
^*\!\psi=\psi-B(r)~,~~~
^*\!r=r~,~~~
^*\!\theta=\theta
\end{eqnarray}
instead of the Kerr-Star coordinate
($T$, $A$ and $B$ are the same functions as (\ref{T}),
(\ref{A}) and (\ref{B})),
the null geodesics are given by
\begin{eqnarray}
\gamma'_{\rm in}&=&
-\frac{\partial}{\partial\, ^*\!r}
+\frac{2}{r^2\Delta_r}V(r)~,
\label{geodesic in:*K}
\\
\gamma'_{\rm out}&=&
\frac{\partial}{\partial\, ^*\!r}~.
\label{geodesic out:*K}
\end{eqnarray}
Then, we obtain a metric
\begin{eqnarray}
g_{BH}&=&
\sum_{i,j=t,\phi,\psi}
g_{ij}d\,^*\!x^id\,^*\!x^j
+g_{\theta\theta}d\theta^2
-2d\, ^*\!t dr
+2a\sin^2\theta drd\,^*\!\phi
+2b\cos^2\theta drd\,^*\psi~,~~
\end{eqnarray}
which differs from (\ref{metric:Kerr-Star})
only in the last three terms.
The Star-Kerr space $^*\!M=\bR_+^2\times$S$^3$
is related to the Kerr-Star space $M^*$
by the following isometric
mapping $f: {}^*\!M\to M^*$;
$ 
^*\!t=-t^*,~
^*\!\phi=-\phi^*,~
^*\!\psi=-\psi^*
$ 
together with
$^*\!r=r^*$ and $^*\!\theta=\theta^*$
(see figure \ref{*K}).

\begin{figure}
   \parbox{\textwidth}{
   \begin{center}
      \psfig{file=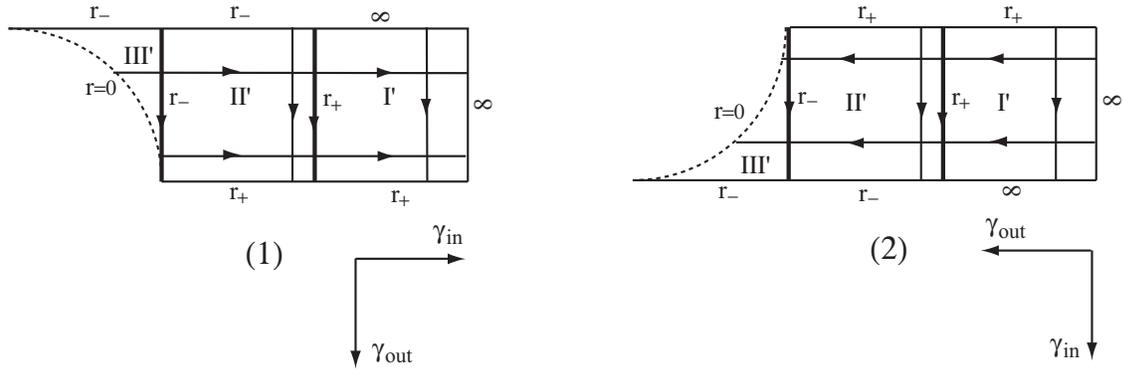,height=50mm}   
   \end{center}
   \caption{\textit{ Kerr-Star$'$ space  and Star-Kerr$'$ space. }~~
   {\small 
   Kerr-Star$'$ space (1) (Star-Kerr$'$ space (2))
   is defined by reversing the directions of
   ingoing and outgoing null geodesics
   of Kerr-Star space (Star-Kerr space, respectively).
   }
   }
\label{K*'}
}
\end{figure}

\begin{figure}
   \parbox{\textwidth}{
   \begin{center}
      \psfig{file=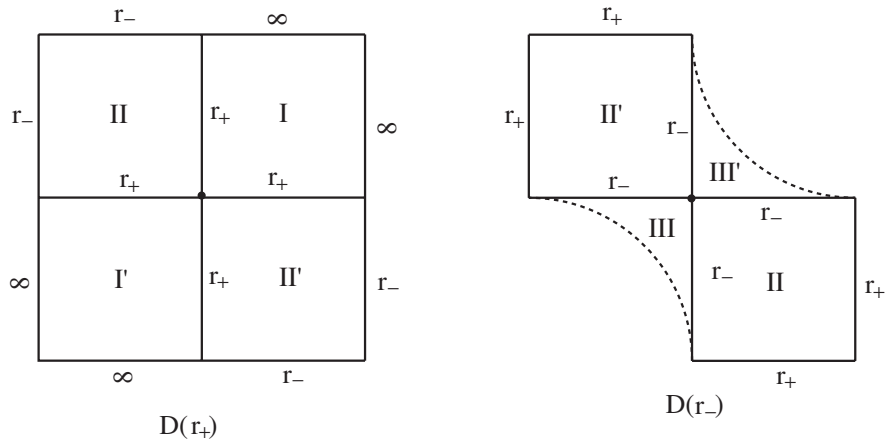,height=60mm}
   \end{center}
   \caption{Kruskal spaces $D(r_\pm)$.}~~
   {\small 
   Kruskal spaces $D(r_\pm)$ are building blocks of the
   maximal extension (see figure 5).
   Dot $\bullet $ at the center represents the crossing three-sphere
   S$^3(r_\pm)$.
   }
\label{Kruskal}
}
\end{figure}

\begin{figure}
   \parbox{\textwidth}{
   \begin{center}
      \psfig{file=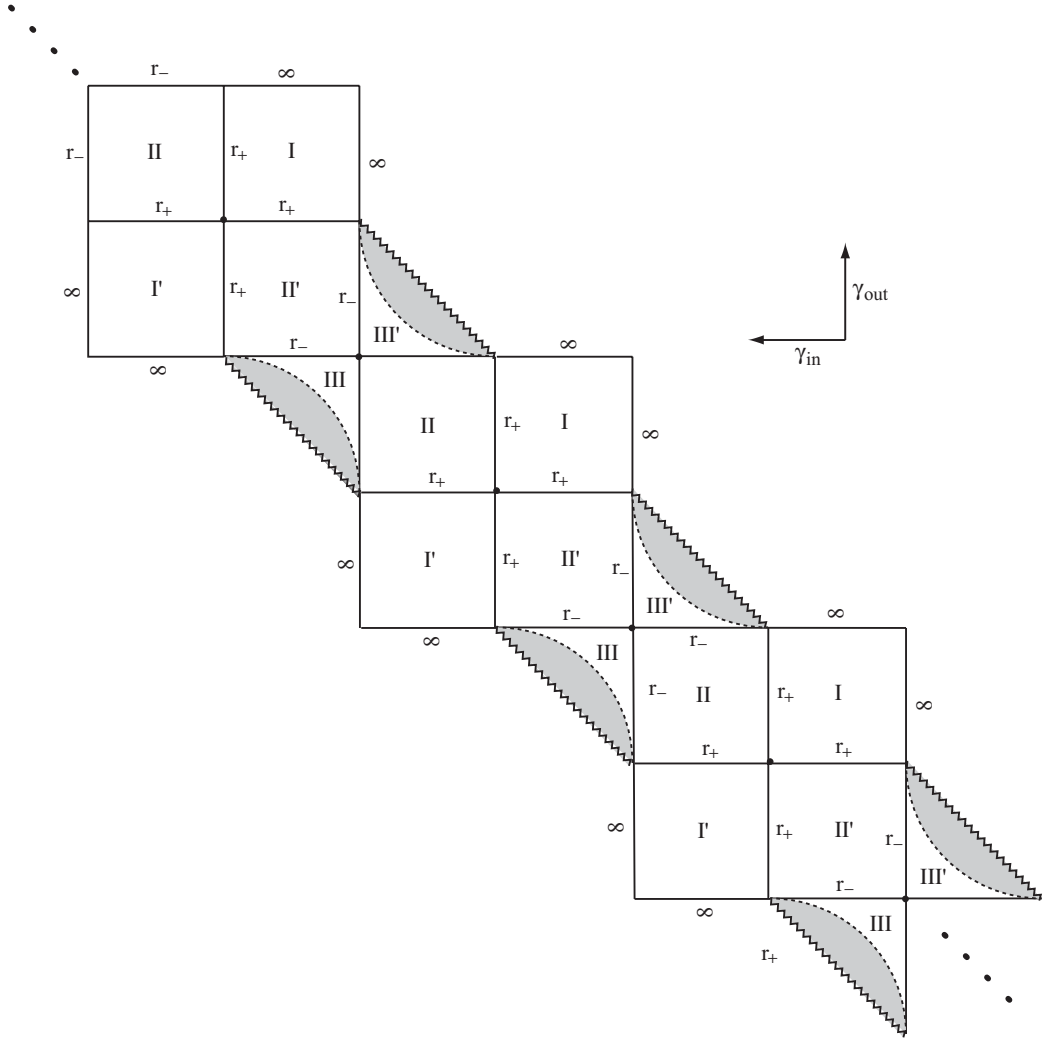,height=140mm}
   \end{center}
   \caption{\textit{Maximal extension.}~~
   {\small 
   Each row corresponds to a Kerr-Star$(')$ space
   while each column a  Star-Kerr$(')$ space.
   The Boyer-Lindquist block III (or III$'$)
   is extended to a negative region
   of the square of radial coordinate $r^2$ \cite{MP},
   which is depicted in gray.
   The wavy lines in III (or III$'$) represent
   curvature singularities defined by
   $r^2=-a^2\sin^2\theta-b^2\cos^2\theta$.
   }
   }
\label{Max}
}
\end{figure}

We may write the black hole metric in terms of the
Kruskal coordinates $(u^{\pm},v^{\pm},\phi^{\pm},\psi^{\pm},\theta)$.
These coordinates are defined on the regions $D(r_{\pm})$ 
(see figures \ref{K*'}, \ref{Kruskal} and \ref{Max});
the relation between Boyer-Lindquist and Kruskal coordinates 
are given by
\begin{eqnarray}
|u^{\pm}| &=& \exp[\kappa_{\pm}(T(r)-t)]~,\\
|v^{\pm}| &=& \exp[\kappa_{\pm}(T(r)+t)]~,\\
\phi^{\pm} &=& \phi-\frac{a t}{r_{\pm}^2+a^2}~,\\
\psi^{\pm} &=& \psi-\frac{b t}{r_{\pm}^2+b^2},
\end{eqnarray}
where the constants
\begin{eqnarray}
\kappa_{\pm} = \frac{r_{\pm}(r_{\pm}^2-r_{\mp}^2)}
{(r_{\pm}^2+a^2)(r_{\pm}^2+b^2)}
\end{eqnarray}
are the surface gravities of the horizons $H_{\pm}$.
The integral (\ref{T}) is calculated as
\begin{eqnarray}
T(r)=r+\frac{1}{2 \kappa_{+}}\log\left|\frac{r-r_{+}}{r+r_{+}}\right|
+\frac{1}{2 \kappa_{-}}\log\left|\frac{r-r_{-}}{r+r_{-}}\right|~.
\end{eqnarray}
Then the radial coordinate $r=r(u^{\pm}v^{\pm})$ is given implicitly
by
\begin{eqnarray}
u^{\pm} v^{\pm}=\frac{1}{G_{\pm}(r)}\left
(\frac{r-r_{\pm}}{r+r_{\pm}} \right),
\end{eqnarray}
where
\begin{eqnarray}
G_{\pm}(r)=\pm\left|\frac{r-r_{\mp}}{r+r_{\mp}}\right|^{r_{\mp}/r_{\pm}}
 \exp(-2 \kappa_{\pm} r)
\end{eqnarray}
are non-zero analytic functions on $D(r_{\pm})$. Note that
two hypersurfaces $u^{\pm}=0$ and $v^{\pm}=0$ express the horizons,
which cross at the region $D(r_{\pm})|_{u_{\pm}=v_{\pm}=0}\simeq S^3(r_{\pm})$
(crossing three-sphere).
The Killing vector fields $V_{\pm}$
on the horizons become
\begin{eqnarray}
V_{\pm}=
\kappa_{\pm}(r_\pm^2+a^2)(r_\pm^2+b^2) 
\left(-u^{\pm}\frac{\partial}{\partial u^{\pm}}
+v^{\pm}\frac{\partial}{\partial v^{\pm}} \right)
\end{eqnarray}
in the Kruskal coordinate. It follows that
the crossing spheres are fixed sets of $V_{\pm}$.

On $D(r_{+})$ the metric takes the form,
\begin{eqnarray}
g_{BH} &=&
\frac{G_{+}^2}{4 \kappa_{+}^2}
\frac{(r+r_{+})^3(r^2-r_{-}^2)}{(r-r_{+})r^2\rho^2} 
\left(
\frac{r^4 \rho^4}{(r^2+a^2)^2(r^2+b^2)^2}-
\frac{r_{+}^4 \rho_{+}^4}{(r_{+}^2+a^2)^2(r_{+}^2+b^2)^2} \right)
\nonumber\\&&~~~~~\times
(u^{+ 2} dv^{+ 2}+v^{+ 2}du^{+ 2}) \nonumber\\
&&+
\frac{G_{+}}{2 \kappa_{+}^2}
\frac{(r+r_{+})^2(r^2-r_{-}^2)}{r^2 \rho^2} \left(
\frac{r^4 \rho^4}{(r^2+a^2)^2(r^2+b^2)^2}+
\frac{r_{+}^4 \rho_{+}^4}{(r_{+}^2+a^2)^2(r_{+}^2+b^2)^2} \right)
du^{+}dv^{+} \nonumber\\
&&+
\frac{G_{+}^2}{4 \kappa_{+}^2} \frac{(r+r_{+})^4(a^2 b^2 \rho_{+}^4+
a^2(r_{+}^2+b^2)^2r^2 \sin^2 \theta+b^2(r_{+}^2+a^2)^2 r^2 \cos^2 \theta)}
{r^2 \rho^2(r_{+}^2+a^2)^2(r_{+}^2+b^2)^2}
\nonumber\\&&~~~~~\times
(u^{+}dv^{+}-v^{+}du^{+})^2 \nonumber\\
&&+
\frac{a G_{+}}{\kappa_{+}}\frac{(r+r_{+})^2(r^2+a^2+\rho_{+}^2)}
{\rho^2(r_{+}^2+a^2)} \sin^2 \theta 
d\phi^{+}(u^{+}dv^{+}-v^{+}du^{+}) \nonumber\\
&&+
\frac{b G_{+}}{\kappa_{+}}\frac{(r+r_{+})^2(r^2+b^2+\rho_{+}^2)}
{\rho^2(r_{+}^2+b^2)} \cos^2 \theta 
d\psi^{+}(u^{+}dv^{+}-v^{+}du^{+})\nonumber\\ 
&&+
g_{S^3(r_{+})}~,
\end{eqnarray}
where $\rho_{+}^2=r_{+}^2+a^2 \cos^2 \theta+b^2 \sin^2 \theta$ and
the last term
\begin{eqnarray}
g_{S^3(r_{+})}=
\sum_{i,j=\phi^{+},\psi^{+}}~ g_{ij} dx^{i} dx^{j}+
\rho^2 d\theta^2~,
\end{eqnarray}
($g_{ij}$ are the same functions as (\ref{p:top})-(\ref{p:bottom}))
gives a metric on the crossing sphere $S^3(r_{+})$. In this
coordinate,
 $S^3(r_{+})$ becomes a totally geodesic submanifold 
of $D(r_{+})$.
The metric on $D(r_{-})$ 
is given by replacing plus indices with minus indices. If we put
$a=b=0$ (Schwarzschild metric),
for which we have $r_{+}=1/\kappa_{+}=\sqrt{2 m}$
and $r_{-}=0$,
then the metric reduces to
\begin{eqnarray}
g_{BH}=2 m \left(1+\frac{\sqrt{2 m}}{r} \right)^2
\exp\left(-\sqrt{\frac{2}{m}}r \right) 
du^{+} dv^{+}+r^2
(d\theta^2+\sin^2 \theta d\phi^2+\cos^2 \theta d\psi^2)~
\nonumber\\
\end{eqnarray}
with
\begin{eqnarray}
u^{+} v^{+}=\left(\frac{r-\sqrt{2m}}{r+\sqrt{2m}} \right)
\exp\left(\sqrt{\frac{2}{m}}r \right).
\end{eqnarray}


\sect{Five-dimensional AdS Kerr black holes}
The analysis
persists in the case with a cosmological constant.
The AdS Kerr metric in the Boyer-Lindquist coordinate
is given as \cite{Hawking}
\begin{eqnarray}
g_{BH}'&=&-\frac{\Delta_r}{\rho^2}
 \left[dt-\frac{a\sin^2\theta}{\Xi_a}d\phi-\frac{b\cos^2\theta}{\Xi_b}d\psi\right]^2
+\frac{\rho^2}{\Delta_r}dr^2
+\frac{\rho^2}{\Delta_\theta}d\theta^2
\nonumber\\&&
+\frac{\Delta_\theta\sin^2\theta}{\rho^2}(adt-\frac{r^2+a^2}{\Xi_a}d\phi)^2
+\frac{\Delta_\theta\cos^2\theta}{\rho^2}(bdt-\frac{r^2+b^2}{\Xi_b}d\psi)^2
\nonumber\\&&
+\frac{1+r^2\ell^2}{r^2\rho^2}\left[
 abdt
 -\frac{b(r^2+a^2)\sin^2\theta}{\Xi_a}d\phi
 -\frac{a(r^2+b^2)\cos^2\theta}{\Xi_b}d\psi
\right]^2~,
\label{AdS Kerr metric}
\end{eqnarray}
where
\begin{eqnarray}
&&
\rho^2=
 r^2
 +a^2\cos^2\theta
 +b^2\sin^2\theta~,
~~~
\Delta_r=
 \frac{1}{r^2}(r^2+a^2)(r^2+b^2)(1+r^2\ell^2)-2m
~,
\nonumber
\\&&
\Delta_\theta=
 1
 -a^2\ell^2\cos^2\theta
 -b^2\ell^2\sin^2\theta~,
~~~
\Xi_a=1-a^2\ell^2~, 
~~~
\Xi_b=1-b^2\ell^2
~.
\label{Delta_r:AdS}
\end{eqnarray}

We find that the theorem 1
is generalized as follows.
\medskip

\noindent
\textbf{\underline{Theorem 2.}}\\
{\it
There exists a constant $K'$ for a geodesic $\gamma$
satisfying
\begin{eqnarray}
&&
{\rho^4}
\left( 
\frac{d\theta}{ds}\right)^2+\Theta_\ell(\theta)=0~,
\\&&
{\rho^4}\left( 
\frac{dr}{ds}\right)^2+R_\ell(r)=0~,
\end{eqnarray}
where
\begin{eqnarray}
\Theta_\ell&=&
(a^2\sin^2\theta+b^2\cos^2\theta-a^2b^2\ell^2)
L_t^2
+\frac{\Xi_a^2(1-b^2\ell^2\sin^2\theta)}{\sin^2\theta}L_\phi^2
+\frac{\Xi_b^2(1-a^2\ell^2\cos^2\theta)}{\cos^2\theta}L_\psi^2
\nonumber\\&&
-2\ell^2\left(
a^2\sin^2\theta+b^2\cos^2\theta-a^2b^2\ell^2
\right)
L_t(aL_\phi+bL_\psi)
-2ab\ell^2\Xi_a\Xi_bL_\phi L_\psi
\nonumber\\&&
-\Delta_\theta(a^2\cos^2\theta+b^2\sin^2\theta) Q
-\Delta_\theta K'~,\\
R_{\ell}&=&
R_{-2}r^{-2}+R_0+R_2r^2+R_4r^4+R_6r^6
~.
\end{eqnarray}
The coefficients are explicitly
given by
\begin{eqnarray}
R_6&=&
-Q\ell^2~,\\
R_4&=&
-L_t^2
+2\ell^2L_t(aL_\phi+bL_\psi)
-(1+a^2\ell^2+b^2\ell^2)Q+\ell^2K'
~,\\
R_2&=&
\left(
-2(a^2+b^2)+a^2b^2\ell^2
\right)
L_t
\left(
L_t
-2\ell^2(aL_\phi+bL_\psi)
\right)
+2ab\ell^2\Xi_a\Xi_bL_\phi L_\psi
\nonumber\\&&
+b^2\ell^2\Xi_a^2L_\phi^2
+a^2\ell^2\Xi_b^2L_\psi^2
-(a^2+b^2+a^2b^2\ell^2-2m)Q
+(1+a^2\ell^2+b^2\ell^2)K'
~,\\
R_0&=&
(-a^4\Xi_b-b^4\Xi_a-3a^2b^2)
L_t^2
+\Xi_a^2(b^2-a^2+a^2b^2\ell^2+b^4\ell^2)L_\phi^2
\nonumber\\&&
+\Xi_b^2(a^2-b^2+a^2b^2\ell^2+a^4\ell^2)L_\psi^2
-2ab\ell^2\Xi_a\Xi_b(a^2+b^2)L_\phi L_\psi
\nonumber\\&&
+2\left(
a^4\ell^2\Xi_b
+b^4\ell^4\Xi_a
+3a^2b^2\ell^2
-2m
\right)
L_t(aL_\phi+bL_\psi)
\nonumber\\&&
-a^2b^2Q
+(a^2+b^2+a^2b^2\ell^2-2m)K'
~,\\
R_{-2}&=&
-a^2b^2(a^2+b^2-a^2b^2\ell^2+2m)L_t^2
-b^2\Xi_a^2(a^2\Xi_b^2-b^2+2m)L_\phi^2
\nonumber\\&&
-a^2\Xi_b^2(b^2\Xi_a^2-a^2+2m)L_\psi^2
+2ab^2
\left(
a^2\ell^2(a^2\Xi_b+b^2)+2m\Xi_a
\right)
L_tL_\phi
+a^2b^2K'
\nonumber\\&&
+2a^2b
\left(
b^2\ell^2(b^2\Xi_a+a^2)+2m\Xi_b
\right)
L_tL_\psi
+2ab\Xi_a\Xi_b(a^2b^2\ell^2-2m)L_\phi L_\psi
~
\end{eqnarray}
with
\begin{eqnarray}
Q=\langle\gamma',\gamma'\rangle~,~~~
L_i=\langle\gamma',\frac{\partial}{\partial x^i}\rangle,~~~
x^i=(t,\phi,\psi).
\end{eqnarray}

}
\medskip

These equations reduce to (\ref{Theta:l=0})
and (\ref{R:l=0}) when we set $\ell=0$.

The special geodesics in
(\ref{special geodesic})
are generalized to
\begin{eqnarray}
\gamma'_{\{\rm{out \atop in }}&=&
\pm\frac{\partial}{\partial r}
+\frac{1}{r^2\Delta_r}V(r)
\label{geodesic:AdS}
\end{eqnarray}
with
\begin{eqnarray}
V(r)&=&
(r^2+a^2)(r^2+b^2)\frac{\partial}{\partial t}
+a\Xi_a(r^2+b^2)\frac{\partial}{\partial \phi}
+b\Xi_b(r^2+a^2)\frac{\partial}{\partial \psi}
~.
\label{V:AdS}
\end{eqnarray}
The constants are
\begin{eqnarray}
&&Q=0~,~~~
L_t=-1~,~~~
L_\phi=\frac{a\sin^2\theta}{\Xi_a}~,~~~
L_\psi=\frac{b\cos^2\theta}{\Xi_b}~,
\nonumber\\&&
K=\frac{2}{\Xi_a\Xi_b}(a^2\sin^2\theta+b^2\cos^2\theta-a^2b^2\ell^2)
~,
\end{eqnarray}
for which we have $\Theta_\ell=0$ and $R_\ell=-\rho^4$.

In the same way as the Ricci flat case, Kerr-Star coordinate and
Star-Kerr coordinate are introduced:
\begin{eqnarray}
{^*\! t}^*=t \pm T(r)~,~~~
^*\!\phi^*=\phi \pm A(r)~,~~~
^*\!\psi^*=\psi \pm B(r)~,~~
\end{eqnarray}
where
\begin{eqnarray}
T(r)&=&
\int \frac{(r^2+a^2)(r^2+b^2)}{r^2\Delta_r}dr~,
\label{T:AdS}\\
A(r)&=&
\int \frac{a\Xi_a(r^2+b^2)}{r^2\Delta_r}dr~,
\label{A:AdS}\\
B(r)&=&
\int \frac{b\Xi_b(r^2+a^2)}{r^2\Delta_r}dr~.
\label{B:AdS}
\end{eqnarray}
Then, the metric is given by
\begin{eqnarray}
g_{BH}'&=&
\sum_{i,j=t,\phi,\psi}
g_{ij}d\,^*\!x^{*i}d\,^*\!x^{*j}
+g_{\theta\theta}d\theta^2
\mp 2d\, ^*\!t^* dr
\pm \frac{2a\sin^2 \theta}{\Xi_a} drd\,^*\!\phi^*
\pm \frac{2b\cos^2 \theta}{\Xi_b} drd\,^*\psi^*~,
\nonumber\\
\end{eqnarray}
where
\begin{eqnarray}
g_{tt}&=&
-1+\frac{2m}{\rho^2}
-\ell^2(
r^2+a^2\sin^2\theta+b^2\cos^2\theta
)
~,
\\
g_{t\phi}&=&
-\frac{2ma\sin^2\theta}{\rho^2\Xi_a}
+\ell^2\frac{a\sin^2\theta}{\Xi_a}(r^2+a^2)~,\\
g_{t\psi}&=&
-\frac{2mb\cos^2\theta}{\rho^2\Xi_b}
+\ell^2\frac{b\cos^2\theta}{\Xi_b}(r^2+b^2)~,\\
g_{\phi\phi}&=&
\sin^2\theta\left[
\frac{r^2+a^2}{\Xi_a}
+\frac{2\sin^2\theta a^2m}{\rho^2\Xi_a^2}
\right]
~,\\
g_{\psi\psi}&=&
\cos^2\theta\left[
\frac{r^2+b^2}{\Xi_b}
+\frac{2\cos^2\theta b^2m}{\rho^2\Xi_b^2}
\right]
~,\\
g_{\phi\psi}&=&
\frac{2abm\cos^2\theta\sin^2\theta}{\rho^2\Xi_a\Xi_b}
~.
\end{eqnarray}
In this coordinate system, 
the special geodesics (\ref{geodesic:AdS})
reduce to
(\ref{geodesic in:*K})
and
(\ref{geodesic out:*K}),
or
(\ref{geodesic in})
and
(\ref{geodesic out}),
with $V$ in (\ref{V:AdS}) and $\Delta_r$ in (\ref{Delta_r:AdS}).

We now describe some curvature property
of the AdS Kerr black hole.
Let us introduce an orthonormal frame $\{e_a\}$ $(a=1,...,5)$:
\begin{eqnarray}
e_1&=&
\frac{1}{r^2\sqrt{\Delta_r\varepsilon}\rho}V~,
\label{e1:AdS}\\
e_2&=&
\frac{\Xi_a\Xi_b}{\Omega_2\sin\theta}\left(
\frac{\hat\Sigma_b}{\Xi_b}W
+\frac{b\hat\Sigma_a}{F}Z
\right)~,\\
e_3&=&
\frac{\Xi_a\Xi_b}{\Omega_3\cos\theta}\left(
\frac{\hat\Sigma_a}{\Xi_a}W
-\frac{a\hat\Sigma_b}{F}Z
\right)~,\\
e_4&=&
\frac{\sqrt{\Delta_r\varepsilon}}{\rho}\frac{\partial}{\partial r}~,\\
e_5&=&
\frac{\sqrt{\Delta_\theta}}{\rho}\frac{\partial}{\partial \theta}~.
\label{e5:AdS}
\end{eqnarray}
The vector field $V$ is given by (\ref{V:AdS}), and
\begin{eqnarray}
W&=&
\frac{a}{\Xi_a}\sin^2\theta\frac{\partial}{\partial t}
+\frac{\partial}{\partial \phi}~,\\
Z&=&
\frac{a}{\Xi_a}\sin^2\theta\frac{\partial}{\partial \psi}
-\frac{b}{\Xi_b}\cos^2\theta\frac{\partial}{\partial \phi}~,
\end{eqnarray}
which satisfy $\langle V,W\rangle=\langle V,Z\rangle=0$.
The functions $\hat\Sigma_a$, $\hat\Sigma_b$
and $F$ are
defined by
\begin{eqnarray}
\hat\Sigma_a=
\sqrt{r^2+\frac{\Delta_\theta}{\Xi_a}a^2}~,~~~
\hat\Sigma_b=
\sqrt{r^2+\frac{\Delta_\theta}{\Xi_b}b^2}~,
\end{eqnarray}
and
\begin{eqnarray}
F&=&
\frac{1}{1-\ell^2\rho^2}\left(
\frac{1}{2}(\Xi_a+\Xi_b)\hat\Sigma_a\hat\Sigma_b
+
\sqrt{\Delta_\theta r^2\rho^2
+\frac{1}{4}\ell^4(a^2-b^2)^2\hat\Sigma_a^2\hat\Sigma_b^2}
\right)~,
\end{eqnarray}
with
\begin{eqnarray}
(\Omega_2)^2&=&
\hat\Sigma_b^2r^2
+(a^2\hat\Sigma_b^2 -b^2\hat\Sigma_a^2)(1-\ell^2\rho^2)\cos^2\theta
-b^2\hat\Sigma_a^3\hat\Sigma_b(\Xi_a-\Xi_b)\frac{\cos^2\theta}{F}~,\\
(\Omega_3)^2&=&
\hat\Sigma_a^2r^2
-(a^2\hat\Sigma_b^2 -b^2\hat\Sigma_a^2)(1-\ell^2\rho^2)\sin^2\theta
+a^2\hat\Sigma_a\hat\Sigma_b^3(\Xi_a-\Xi_b)\frac{\sin^2\theta}{F}~.
\end{eqnarray}
If we put $\ell=0$,
the equations (\ref{e1:AdS})-(\ref{e5:AdS}) 
reduce to (\ref{e1:MP})-(\ref{e5:MP}).
We consider the Weyl curvature $W^{ab}{}_{cd}$
as a linear map on two forms.
The corresponding matrix $\hat W^{ab}{}_{cd}$
defined by  $W^{ab}{}_{cd}=\frac{2m}{\rho^6}\hat W^{ab}{}_{cd}$ 
takes the form (\ref{matrix}).
Note that the functions {\rm\bf I} and {\rm\bf J}
defined in (\ref{I and J})
are replaced with $\hat {\rm\bf I}$ and  $\hat {\rm\bf J}$,
respectively, as\fnote{$\sharp $}{
We have explicitly checked these formulae by using Maple.
}
\begin{eqnarray}
\hat {\rm\bf I}&=&
\frac{a\cos\theta}{r}
\frac{(\hat\Sigma_b F-b^2\hat\Sigma_a)\sqrt{\Delta_\theta}\rho}{F\Omega_2}~,
~~~
\hat {\rm\bf J}=
\frac{b\sin\theta}{r}
\frac{(\hat\Sigma_a F-a^2\hat\Sigma_b)\sqrt{\Delta_\theta}\rho}{F\Omega_3}~.
\end{eqnarray}
In general, they depend on the cosmological constant $\ell$,
but the combination $\hat{\rm\bf K}^2\equiv \hat {\rm\bf I}^2+\hat {\rm\bf J}^2$
coincides with ${\rm\bf K}^2={\rm\bf I}^2+{\rm\bf J}^2$,
i.e.
$\hat {\rm\bf K}^2={\rm\bf K}^2$.
As special cases, we have $\hat {\rm\bf I}={\rm\bf I}$
and $\hat {\rm\bf J}={\rm\bf J}$
for $a=b$ or $a\neq 0, b=0$ ($a=0, b\neq 0$).
We find that the eigenvalues are exactly the same 
as (\ref{lambda 1})-(\ref{lambda 6}).
Thus we state as follows.
\medskip

\noindent
{\bf \underline{Theorem 3.}}\\
{\it
Five-dimensional AdS Kerr black holes are
isospectrum deformations of Ricci-flat
Kerr black holes in the sense that
the eigenvalues of the Weyl curvature
are preserved.
}

\medskip

\noindent
\underline{Remark 1.}\\
{\it 
We conjecture that the statement above
is true for the general AdS Kerr black holes in
all dimensions \cite{Gibbons}
(see appendix A for AdS Kerr black holes in
four-dimensions).
}

\medskip

\noindent
\underline{Remark 2.}\\
{\it
Changing the negative cosmological constant
to the positive one, $\ell^2\to -\ell^2$,
we obtain the same results.
}

\medskip

Finally we discuss the relation
between the Weyl curvature of
AdS Kerr black holes with equal angular momenta
$a=b$
and that of Sasaki-Einstein metrics constructed in \cite{GMSW:5}.
According to  \cite{CLP} \cite{HSY2}, we write
the five-dimensional AdS black hole metric with a
negative cosmological constant $4\Lambda$
as
\begin{eqnarray}
g&=&-\frac{W(R)}{b(R)}dt^2
+\frac{dR^2}{W(R)}
+R^2\left(\frac{1}{4}(d\theta^2+\sin^2\theta d\phi^2)
+
b(R)\left(d\psi +\frac{1}{2}\cos\theta d\phi
+f(R)dt\right)^2 \right)
,\nonumber\\&&
\label{metric: twist}
\end{eqnarray}
where
\begin{eqnarray}
W&=&
1
-\Lambda R^2
-\frac{2m(\delta^2+\Lambda J^2)}{R^2}
+\frac{2mJ^2}{R^4}~,\nonumber\\
b&=&
1
+\frac{2mJ^2}{R^4}~,~~~
f=\frac{1}{J}\left(
1-\frac{\delta}{b}
\right)~.
\end{eqnarray}
The metric is parameterized by the mass $m$,
the angular momentum $J$
and an unphysical parameter $\delta$ \cite{CLP}.
The eigenvalues of the Weyl curvature
are calculated in the same way;
\begin{eqnarray}
2&~:~&\lambda_1=
\frac{2m}{R^4}(\delta^2+\Lambda J^2)~,
\label{lambda 1:twist}\\
2&~:~&\lambda_2=-
\frac{2m}{R^4}(\delta^2+\Lambda J^2)~,\\
2&~:~&\lambda_3=
\frac{2m}{R^6}\left(
2J^2-(\delta^2+\Lambda J^2)R^2
+2J\sqrt{J^2-(\delta^2+\Lambda J^2)R^2)}
\right)~,\\
2&~:~&\lambda_4=
\frac{2m}{R^6}\left(
2J^2-(\delta^2+\Lambda J^2)R^2
-2J\sqrt{J^2-(\delta^2+\Lambda J^2)R^2)}
\right)~,\\
1&~:~&\lambda_5=
\frac{2m}{R^6}\left(
-4J^2+2(\delta^2+\Lambda J^2)R^2
-\sqrt{16J^4-16J^2(\delta^2+\Lambda J^2)R^2+(\delta^2+\Lambda J^2)^2R^4}
\right)~,\nonumber\\&&\\
1&~:~&\lambda_6=
\frac{2m}{R^6}\left(
-4J^2+2(\delta^2+\Lambda J^2)R^2
+\sqrt{16J^4-16J^2(\delta^2+\Lambda J^2)R^2+(\delta^2+\Lambda J^2)^2R^4}
\right)~,\nonumber\\&&
\label{lambda 6: twist}
\end{eqnarray}
where the number in the left hand side represents
the degeneracy.
If we introduce a new radial coordinate $r$ defined by
\begin{eqnarray}
R^2=\sqrt{\delta^2+\Lambda J^2}\left(
r^2+\frac{J^2}{(\delta^2+\Lambda J^2)^{3/2}}
\right),
\label{R and r}
\end{eqnarray}
then the eigenvalues reduce to
(\ref{lambda 1})-(\ref{lambda 6})
with
\begin{eqnarray}
a^2=b^2=\frac{J^2}{(\delta^2+\Lambda J^2)^{3/2}}~.
\label{parameter}
\end{eqnarray}
The Sasaki-Einstein metric
appears by setting $\delta^2+\Lambda J^2=0$
(together with an Wick rotation)
as shown in \cite{HSY2}.
Then the equations (\ref{lambda 1:twist})-(\ref{lambda 6: twist})
yield
\begin{eqnarray}
\lambda_1=\lambda_2=\lambda_4=\lambda_6=0~,~~~
\lambda_3=\frac{8mJ^2}{R^6}~,~~~
\lambda_5=-\frac{16mJ^2}{R^6}~,
\end{eqnarray}
so that the multiplicity changes at this special value of $\delta$.
Note that from (\ref{R and r}) and (\ref{parameter})
this setting is not allowed
in the parameterization of the metric (\ref{AdS Kerr metric}).

\bigskip

\section*{Acknowledgements}
The authors thank
Yoshitake Hashimoto,
Hideki Ishihara and Ken-ichi Nakao
for useful discussions.
We also thank Gary Gibbons
for correspondence on the theorem 3.
This paper is supported by the 21 COE program
``Constitution of wide-angle mathematical basis focused on knots".
Research of Y.Y. is supported  in part by the Grant-in
Aid for scientific Research (No.~14540073 and No.~14540275)
from Japan Ministry of Education.
The preliminary version of this work 
was presented by Y.Y.
in ``Quantum Cohomology and Mirror Symmetry Day''
held at Tokyo Metropolitan University
(21 January, 2005).

\appendix
\sect{Four-dimensional AdS Kerr black holes}
In this appendix, we briefly describe 
geodesics and the Weyl curvature of
four-dimensional AdS Kerr black holes.

The metric with a negative cosmological constant $-3\ell^2$ is given as
\begin{eqnarray}
g_{BH}^{(4)}=
-\frac{\Delta_r}{\rho^2}
\left(dt-\frac{a}{\Xi}\sin^2\theta d\phi\right)^2
+\frac{\rho^2}{\Delta_r}dr^2
+\frac{\rho^2}{\Delta_\theta}d\theta^2
+\frac{\sin^2\theta \Delta_\theta}{\rho^2}
 \left(a dt-\frac{r^2+a^2}{\Xi}d\phi\right)^2
,~~~~~~
\end{eqnarray}
where~~~~~
\begin{eqnarray}
&&\rho^2=r^2+a^2\cos^2\theta~,~~~
\Delta_r=(r^2+a^2)(1+\ell^2r^2)-2mr~,
\nonumber\\&&
\Delta_\theta=1-\ell^2a^2\cos^2\theta~,~~~
\Xi=1-\ell^2a^2~,
\end{eqnarray}
which is parameterized by the mass $m$ and the angular momentum $a$.
The orthonormal frame $\{e_a\}$ $(a=1,2,3,4)$ is given as
\begin{eqnarray}
e_1&=&
\frac{1}{\sqrt{\Delta_r}\rho}\left(
(r^2+a^2)\frac{\partial}{\partial t}
+a\Xi \frac{\partial}{\partial \phi}
\right)~,\\
e_2&=&
\frac{1}{\sqrt{\Delta_\theta}\rho\sin\theta}
\left(
a\sin^2\theta\frac{\partial}{\partial t}
+\Xi\frac{\partial}{\partial \phi}
\right)~,\\
e_3&=&\frac{\sqrt{\Delta_r}}{\rho}\frac{\partial}{\partial r}~,\\
e_4&=&
\frac{\sqrt{\Delta_\theta}}{\rho}\frac{\partial}{\partial \theta}~.
\end{eqnarray}

The four-dimensional version
of the theorem 2 states as follows.
{\it
There exists a constant $K'$
for a geodesic $\gamma$ satisfying
\begin{eqnarray}
\rho^4\left(
\frac{d\theta}{ds}
\right)^2
+\Theta_\ell(\theta)=0~,~~~
\rho^4\left(
\frac{dr}{ds}
\right)^2
+R_\ell(r)=0~,
\end{eqnarray}
where
\begin{eqnarray}
\Theta_\ell&=&
a^2\sin^2\theta L_t^2
+\frac{\Xi^2}{\sin^2\theta}L_\phi^2
+2a\Xi L_tL_\phi
-a^2\cos^2\theta\Delta_\theta Q
-\Delta_\theta K'~,\\
R_\ell&=&
R_0
+R_1r
+R_2r^2
+R_3r^3
+R_4r^4
+R_6r^6~.
\end{eqnarray}
The coefficients are explicitly given as
\begin{eqnarray}
R_6&=&-Q\ell^2~,\\
R_4&=&-L_t^2-(1+a^2\ell^2)Q+\ell^2 K'~,\\
R_3&=&2m~,\\
R_2&=&
-2a^2L_t^2
-2a\Xi L_tL_\phi
-a^2Q
+(1+a^a\ell^2)K'~,\\
R_1&=&-2mK'~,\\
R_0&=&
a^2(-a^2L_t^2-\Xi^2L_\phi^2-2a\Xi L_tL_\phi+K')~,
\end{eqnarray}
with
$
Q=\langle\gamma',\gamma'\rangle,~
L_i=\langle\gamma',\frac{\partial}{\partial x^i}\rangle,~
x^i=(t, \phi)
$.
}

Next, we consider a linear map on two-forms
defined by the Weyl curvature $W^{ab}{}_{cd}$.
We find that the non-zero components of
the matrix $\hat W^{ab}{}_{cd}$
with $W^{ab}{}_{cd}=\frac{2m}{\rho^2}\hat W^{ab}{}_{cd}$
are given as
\begin{eqnarray}
  \begin{array}{c|cccccc}
  (a,b)\setminus (c,d)     &
  (1,2)    &(1,3)    &(1,4)    &(2,3)    &(2,4)    &(4,5)    \\\hline
  (1,2)     &
  -{\rm\bf I}    &    &    &    &    & -{\rm\bf J}   \\
  (1,3)     &
      &2{\rm\bf I}    &    &    & -2{\rm\bf J}   &    \\
  (1,4)    &
      &    & -{\rm\bf I}   &-{\rm\bf J}    &    &    \\
  (2,3)     &
      &    &{\rm\bf J}    &-{\rm\bf I}    &    &    \\
  (2,4)     &
      & 2{\rm\bf J}   &    &    &2{\rm\bf I}    &    \\
  (3,4)     &
  {\rm\bf J}    &    &    &    &    &-{\rm\bf I}    \\
  \end{array}
\end{eqnarray}
where ${\rm\bf I}$ and ${\rm\bf J}$ are defined as
\begin{eqnarray}
{\rm\bf I}=\frac{1}{2}r(r^2-3a^2\cos^2\theta)~,~~~
{\rm\bf J}=\frac{1}{2}a\cos\theta(3r^2-a^2\cos^2\theta)~.
\end{eqnarray}
It should be noted that this matrix is independent of $\ell$,
and the same as the one examined in \cite{O'Neill}
for the Ricci-flat Kerr black hole.
The eigenvalues are
\begin{eqnarray}
2&~:~&\lambda_1=-{\rm\bf I}+i{\rm\bf J}~,\\
2&~:~&\lambda_2=-{\rm\bf I}-i{\rm\bf J}~,\\
1&~:~&\lambda_3=2{\rm\bf I}+2i{\rm\bf J}~,\\
1&~:~&\lambda_4=2{\rm\bf I}-2i{\rm\bf J}~,
\end{eqnarray}
where the number in the left hand side represents
the degeneracy.
Thus, the theorem 3
persists in four-dimensions;
\textit{four-dimensional AdS Kerr black holes
are
isospectrum deformations of Ricci-flat
Kerr black holes.}


\end{document}